\documentclass[12pt]{article}
\usepackage{epsfig}
\usepackage{amsmath}
\usepackage{hhline}
\usepackage{amssymb}
\usepackage{times}
\usepackage{cite}

\newlength{\dinwidth}
\newlength{\dinmargin}
\begin{document}
\newcommand{\pom}{{I\!\!P}}
\newcommand{\reg}{{I\!\!R}}
\newcommand{\slowpi}{\pi_{\mathit{slow}}}
\newcommand{\fiidiii}{F_2^{D(3)}}
\newcommand{\fiidiiiarg}{\fiidiii\,(\beta,\,Q^2,\,x)}
\newcommand{\n}{1.19\pm 0.06 (stat.) \pm0.07 (syst.)}
\newcommand{\nz}{1.30\pm 0.08 (stat.)^{+0.08}_{-0.14} (syst.)}
\newcommand{\fiidiiiful}{F_2^{D(4)}\,(\beta,\,Q^2,\,x,\,t)}
\newcommand{\fiipom}{\tilde F_2^D}
\newcommand{\ALPHA}{1.10\pm0.03 (stat.) \pm0.04 (syst.)}
\newcommand{\ALPHAZ}{1.15\pm0.04 (stat.)^{+0.04}_{-0.07} (syst.)}
\newcommand{\fiipomarg}{\fiipom\,(\beta,\,Q^2)}
\newcommand{\pomflux}{f_{\pom / p}}
\newcommand{\nxpom}{1.19\pm 0.06 (stat.) \pm0.07 (syst.)}
\newcommand {\gapprox}
   {\raisebox{-0.7ex}{$\stackrel {\textstyle>}{\sim}$}}
\newcommand {\lapprox}
   {\raisebox{-0.7ex}{$\stackrel {\textstyle<}{\sim}$}}
\def\gsim{\,\lower.25ex\hbox{$\scriptstyle\sim$}\kern-1.30ex%
\raise 0.55ex\hbox{$\scriptstyle >$}\,}
\def\lsim{\,\lower.25ex\hbox{$\scriptstyle\sim$}\kern-1.30ex%
\raise 0.55ex\hbox{$\scriptstyle <$}\,}
\newcommand{\pomfluxarg}{f_{\pom / p}\,(x_\pom)}
\newcommand{\dsf}{\mbox{$F_2^{D(3)}$}}
\newcommand{\dsfva}{\mbox{$F_2^{D(3)}(\beta,Q^2,x_{I\!\!P})$}}
\newcommand{\dsfvb}{\mbox{$F_2^{D(3)}(\beta,Q^2,x)$}}
\newcommand{\dsfpom}{$F_2^{I\!\!P}$}
\newcommand{\gap}{\stackrel{>}{\sim}}
\newcommand{\lap}{\stackrel{<}{\sim}}
\newcommand{\fem}{$F_2^{em}$}
\newcommand{\tsnmp}{$\tilde{\sigma}_{NC}(e^{\mp})$}
\newcommand{\tsnm}{$\tilde{\sigma}_{NC}(e^-)$}
\newcommand{\tsnp}{$\tilde{\sigma}_{NC}(e^+)$}
\newcommand{\st}{$\star$}
\newcommand{\sst}{$\star \star$}
\newcommand{\ssst}{$\star \star \star$}
\newcommand{\sssst}{$\star \star \star \star$}
\newcommand{\tw}{\theta_W}
\newcommand{\sw}{\sin{\theta_W}}
\newcommand{\cw}{\cos{\theta_W}}
\newcommand{\sww}{\sin^2{\theta_W}}
\newcommand{\cww}{\cos^2{\theta_W}}
\newcommand{\trm}{m_{\perp}}
\newcommand{\trp}{p_{\perp}}
\newcommand{\trmm}{m_{\perp}^2}
\newcommand{\trpp}{p_{\perp}^2}
\newcommand{\alp}{\alpha_s}

\newcommand{\alps}{\alpha_s}
\newcommand{\sqrts}{$\sqrt{s}$}
\newcommand{\LO}{$O(\alpha_s^0)$}
\newcommand{\Oa}{$O(\alpha_s)$}
\newcommand{\Oaa}{$O(\alpha_s^2)$}
\newcommand{\PT}{p_{\perp}}
\newcommand{\JPSI}{J/\psi}
\newcommand{\sh}{\hat{s}}
\newcommand{\uh}{\hat{u}}
\newcommand{\MP}{m_{J/\psi}}
\newcommand{\PO}{I\!\!P}
\newcommand{\xbj}{x}
\newcommand{\xpom}{x_{\PO}}
\newcommand{\ttbs}{\char'134}
\newcommand{\xpomlo}{3\times10^{-4}}
\newcommand{\xpomup}{0.05}
\newcommand{\dgr}{^\circ}
\newcommand{\pbarnt}{\,\mbox{{\rm pb$^{-1}$}}}
\newcommand{\gev}{\,\mbox{GeV}}
\newcommand{\WBoson}{\mbox{$W$}}
\newcommand{\fbarn}{\,\mbox{{\rm fb}}}
\newcommand{\fbarnt}{\,\mbox{{\rm fb$^{-1}$}}}
%
%
\newcommand{\qsq}{\ensuremath{Q^2} }
\newcommand{\gevsq}{\ensuremath{\mathrm{GeV}^2} }
\newcommand{\et}{\ensuremath{E_t^*} }
\newcommand{\rap}{\ensuremath{\eta^*} }
\newcommand{\gp}{\ensuremath{\gamma^*}p }
\newcommand{\dsiget}{\ensuremath{{\rm d}\sigma_{ep}/{\rm d}E_t^*} }
\newcommand{\dsigrap}{\ensuremath{{\rm d}\sigma_{ep}/{\rm d}\eta^*} }
\def\Journal#1#2#3#4{{#1} {\bf #2} (#3) #4}
\def\NCA{\em Nuovo Cimento}
\def\NIM{\em Nucl. Instrum. Methods}
\def\NIMA{{\em Nucl. Instrum. Methods} {\bf A}}
\def\NPB{{\em Nucl. Phys.}   {\bf B}}
\def\PLB{{\em Phys. Lett.}   {\bf B}}
\def\PRL{\em Phys. Rev. Lett.}
\def\PRD{{\em Phys. Rev.}    {\bf D}}
\def\ZPC{{\em Z. Phys.}      {\bf C}}
\def\EJC{{\em Eur. Phys. J.} {\bf C}}
\def\CPC{ Comp. Phys. Commun.}

\newcommand{\etm}{\ensuremath{E_t^{miss}}}
\newcommand{\eega}{$e^* \rightarrow e \gamma$~}
\newcommand{\enwqq}{$e^* \rightarrow \nu W_{\hookrightarrow q \bar{q}} $~}
\newcommand{\eezqq}{$e^* \rightarrow e Z_{\hookrightarrow q \bar{q}} $~}
\newcommand{\md}{M_{jj}}

\begin{titlepage}
\noindent
DESY 02-096  \hfill  ISSN 0418-9833 \\
July 2002

\vspace*{3cm}

\begin{center}
\begin{Large}
  {\bf
        Search for Excited Electrons at HERA}\\

\vspace{2cm}

H1 Collaboration

\end{Large}
\end{center}

\vspace{2cm}

\begin{abstract}
\noindent A search for excited electron ($e^*$) production is
described in which the electroweak decays \eega, $e^* \rightarrow
e Z$ and $e^* \rightarrow \nu W$ are considered. The data used
correspond to an integrated luminosity of $120$~pb$^{-1}$ taken in
$e^{\pm}p$ collisions from 1994 to 2000 with the H1 detector at
HERA at centre-of-mass energies of 300 and~318 GeV. No evidence
for a signal is found. Mass dependent exclusion limits are derived
for the ratio of the couplings to the compositeness scale,
$f/\Lambda$. These limits extend the excluded region to higher
masses than has been possible in previous direct searches for
excited electrons.
\end{abstract}
\vspace{1.5cm}

\begin{center}
To be submitted to \\
Physics Letters B
\end{center}

\end{titlepage}

%
%

\begin{flushleft}

C.~Adloff$^{33}$,              
V.~Andreev$^{24}$,             
B.~Andrieu$^{27}$,             
T.~Anthonis$^{4}$,             
A.~Astvatsatourov$^{35}$,      
A.~Babaev$^{23}$,              
J.~B\"ahr$^{35}$,              
P.~Baranov$^{24}$,             
E.~Barrelet$^{28}$,            
W.~Bartel$^{10}$
% ...

\bigskip{\it
 $ ^{1}$ I. Physikalisches Institut der RWTH, Aachen, Germany$^{ a}$ \\
% ...
$ ^{4}$ Inter-University Institute for High Energies ULB-VUB, Brussels;
          Universiteit Antwerpen (UIA), Antwerpen; Belgium$^{ c}$ \\
% ...
$ ^{10}$ DESY, Hamburg, Germany \\
% ...
$ ^{23}$ Institute for Theoretical and Experimental Physics,
          Moscow, Russia$^{ l}$ \\
 $ ^{24}$ Lebedev Physical Institute, Moscow, Russia$^{ e}$ \\
% ...
$ ^{27}$ LPNHE, Ecole Polytechnique, IN2P3-CNRS, Palaiseau, France \\
 $ ^{28}$ LPNHE, Universit\'{e}s Paris VI and VII, IN2P3-CNRS,
          Paris, France \\
% ...
$ ^{33}$ Fachbereich Physik, Bergische Universit\"at Gesamthochschule
          Wuppertal, Wuppertal, Germany \\
% ...
$ ^{35}$ DESY, Zeuthen, Germany \\
% ...
$ ^a$ Supported by the Bundesministerium f\"ur Bildung und Forschung, FRG,
      under contract numbers 05 H1 1GUA /1, 05 H1 1PAA /1, 05 H1 1PAB /9,
      05 H1 1PEA /6, 05 H1 1VHA /7 and 05 H1 1VHB /5 \\
% ...
$ ^c$ Supported by FNRS-FWO-Vlaanderen, IISN-IIKW and IWT \\
% ...
$ ^e$ Supported by the Deutsche Forschungsgemeinschaft \\
% ...
$ ^l$ Partially Supported by Russian Foundation
      for Basic Research, grant    no. 00-15-96584 \\
}

\end{flushleft}

\newpage

Among the unexplained features of the Standard Model (SM) of
particle physics is the existence of three distinct generations of
fermions and the hierarchy of their masses. One possible
explanation for this is fermion substructure, with the
constituents of the known fermions being strongly bound together
by a new, as yet undiscovered
force~\cite{Harari:1984xy,Boudjema:1993em}. A natural consequence
of these models would be the existence of excited states of the
known leptons and quarks. Assuming a compositeness scale in
the~TeV region, one would naively expect that the excited fermions
have masses in the same energy region. However, the dynamics of
the constituent level are unknown, so the lowest excited states
could have masses of the order of only a few hundred~${\rm GeV}$.
Electron\footnote{The term ``electron'' is used generically to
refer to both electrons and positrons.}-proton interactions at
very high energies provide an excellent environment in which to
search for excited fermions of the first generation. These excited
electrons ($e^*$) could be singly produced through $t$-channel
$\gamma $ and $Z$ boson exchange. Their production cross-section
and partial decay widths have been calculated using an effective
Lagrangian~\cite{Hagiwara:1985wt,Baur:1990kv} which depends on a
compositeness mass scale $\Lambda$ and on weight factors $f$ and
$f'$ describing the relative coupling strengths of the excited
lepton to the  $SU(2)_L$ and $U(1)_Y$ gauge bosons, respectively.
In this model the excited electron can decay to an electron or a
neutrino via the radiation of a gauge boson ($\gamma$, $W$, $Z$)
with branching ratios determined by the $e^*$ mass and the
coupling parameters $f$ and $f'$. In most
analyses~\cite{OPAL,Abreu:1998jw,ZEUS} the assumption is made that
these coupling parameters are of comparable strength and only the
relationships $f=+f'$ or $f=-f'$ are considered. If a relationship
between $f$ and $f'$ is assumed, the production cross-section and
partial decay widths depend on two parameters only, namely the
$e^*$ mass and the ratio $f/\Lambda$.

In this paper excited electrons are searched for in three samples
of data taken by the H1 experiment from 1994 to 2000 with a total
integrated luminosity of 120 pb$^{-1}$. The first sample consists
of $e^+p$ data accumulated from 1994 to 1997 at positron and
proton beam energies of 27.5~${\rm GeV}$ and 820~${\rm GeV}$
respectively, and corresponds to an integrated luminosity of 37
pb$^{-1}$. A search for excited electrons using this sample of
data has been previously published~\cite{Adloff:2000gv}. The
strategy for the selection of events has been modified from the
procedures described in~\cite{Adloff:2000gv} to optimize the
sensitivity to higher $e^*$ masses. The two other samples were
taken from 1998 to 2000 with an electron or positron beam energy
of 27.5~${\rm GeV}$ and a proton beam energy of 920~${\rm GeV}$.
The integrated luminosities of the $e^- p$ and  $e^+ p$ samples
are 15 pb$^{-1}$ and 68 pb$^{-1}$, respectively. Compared to
previous H1 results~\cite{Adloff:2000gv} the analysis presented
here benefits from an increase in luminosity by a factor of more
than three and an increase of the centre-of-mass energy from 300
GeV to 318 GeV.

We search for all electroweak decays \eega, $e^* \rightarrow e Z$
and $e^* \rightarrow \nu W$, considering the subsequent $Z$ and
$W$ hadronic decay modes only. This leads to final states
containing an electron and a photon, an electron and jets or jets
with missing transverse energy induced by the neutrinos escaping
from the detector. The Standard Model backgrounds which could
mimic such signatures are neutral current Deep Inelastic
Scattering (NC DIS), charged current Deep Inelastic Scattering (CC
DIS), QED Compton scattering (or Wide Angle Bremsstrahlung WAB),
photoproduction processes ($\gamma p$) and lepton pair production
via the two photon fusion process ($\gamma \gamma$).

The determination of the contribution of NC DIS processes is
performed using two Monte Carlo samples which employ different
models of QCD radiation. The first was produced with the
DJANGO~\cite{Schuler:1991yg} event generator which includes QED
first order radiative corrections based on
HERACLES~\cite{Kwiatkowski:1992es}. QCD radiation is implemented
using ARIADNE~\cite{aria} based on the Colour Dipole
Model~\cite{Andersson:1989gp}. This sample, with an integrated
luminosity of more than 10 times the experimental luminosity, is
chosen to estimate the NC DIS contribution in the \eega analysis.
The second sample was generated with the program
RAPGAP~\cite{rapg}, in which QED first order radiative corrections
are implemented as described above. RAPGAP includes the leading
order QCD matrix element and higher order radiative corrections
are modelled by leading-log parton showers. This sample is used to
determine potential NC DIS background in the \enwqq and \eezqq
searches, as RAPGAP describes better this particular phase space
domain~\cite{Adloff:2000gv}. For both samples the parton densities
in the proton are taken from the MRST~\cite{Martin:1998sq}
parametrization which includes constraints from DIS measurements
at HERA up to squared momentum transfers $Q^2 =$ 5000 ${\rm
GeV^2}$~\cite{Aid:1996au,Adloff:1997mf,Derrick:1996ef,Derrick:1996hn}.
Hadronisation is performed in the Lund string fragmentation scheme
using  JETSET~\cite{Sjostrand:1995iq}. The modelling of the CC DIS
process is performed using the  DJANGO program with MRST structure
functions. While inelastic WAB is treated using the DJANGO
generator, elastic and quasi-elastic WAB is simulated with the
WABGEN~\cite{wabgen} event generator. Direct and resolved $\gamma
p$ processes including prompt photon production are generated with
the PYTHIA~\cite{Sjostrand:1994yb} event generator. Finally the
$\gamma \gamma$ process is produced using the LPAIR
generator~\cite{lpair}.

For the calculation of the $e^*$ production cross-section and to
determine the efficiencies, events have been generated with the
COMPOS~\cite{Kohler:1991yu} generator based on the cross-section
formulae given in reference~\cite{Hagiwara:1985wt} and the partial
decay widths stated in reference~\cite{Baur:1990kv}. Initial state
radiation of a photon from the incoming electron is included. This
generator uses the narrow width approximation (NWA) for the
calculation of the production cross section and takes into account
the natural width for the $e^*$ decay. For all values of
$f/\Lambda$ relevant to this analysis this assumption is valid
even at high $e^*$ masses where the natural width of the $e^*$ is
of the order of the experimental resolution. To give an example,
for $M_{e^*}=250$~GeV this resolution is equal to 7~GeV, 10~GeV,
and 12~GeV for the $e \gamma$, $e Z$ and $\nu W$ decay modes,
respectively. All Monte Carlo samples are subjected to a detailed
simulation of the response of the H1 detector.

The detector components of the H1 experiment~\cite{Abt:1997hi1}
most relevant for this analysis are briefly described in the
following. Surrounding the interaction region is a system of drift
and proportional chambers which covers the polar
angle\footnote{The polar angle $\theta$ is measured with respect
to the proton beam direction ($+z$).}  range $7^{\circ} < \theta <
176^{\circ}$. The tracking system is surrounded by a finely
segmented liquid argon (LAr) calorimeter covering the polar angle
range 4$^{\circ}~<~\theta~<154^{\circ}$~\cite{Andrieu:1993kh} with
energy resolutions of
 $\sigma_E / E \simeq 12\% / \sqrt{E} \oplus 1\%$ for
 electrons and $ \sigma_E / E \simeq 50\% / \sqrt{E} \oplus 2\%$ for
 hadrons, obtained in test beam measurements
~\cite{Andrieu:1994yn,Andrieu}. The tracking system and
calorimeters are surrounded by a superconducting solenoid and an
iron yoke instrumented with streamer tubes. Backgrounds not
related to $e^+ p$ or $e^- p$ collisions are rejected by requiring
that a primary interaction vertex be reconstructed within $\pm
35$~cm of the nominal vertex position, by using filters based on
the event topology and by requiring an event time which is
consistent with the interaction time. Electromagnetic clusters are
required to have more than 95\% of their energy in the
electromagnetic part of the calorimeter and to be isolated from
other particles~\cite{Schoening}. They are further differentiated
into electron and photon candidates using the tracking chambers.
Jets with a minimum transverse momentum of 5~GeV are reconstructed
from the hadronic final state using a cone algorithm adapted from
the LUCELL scheme in the JETSET package~\cite{Sjostrand:1995iq}.
Missing transverse energy (\etm) is reconstructed using the vector
sum of energy depositions in the calorimeter cells. The analysis
presented in this paper is described extensively
in~\cite{Nicolas}.

\begin{figure}[hhhh]
\begin{center}

\hspace*{-1.5cm}\begin{tabular}{cc}

\epsfxsize=0.4\textwidth
 \epsffile{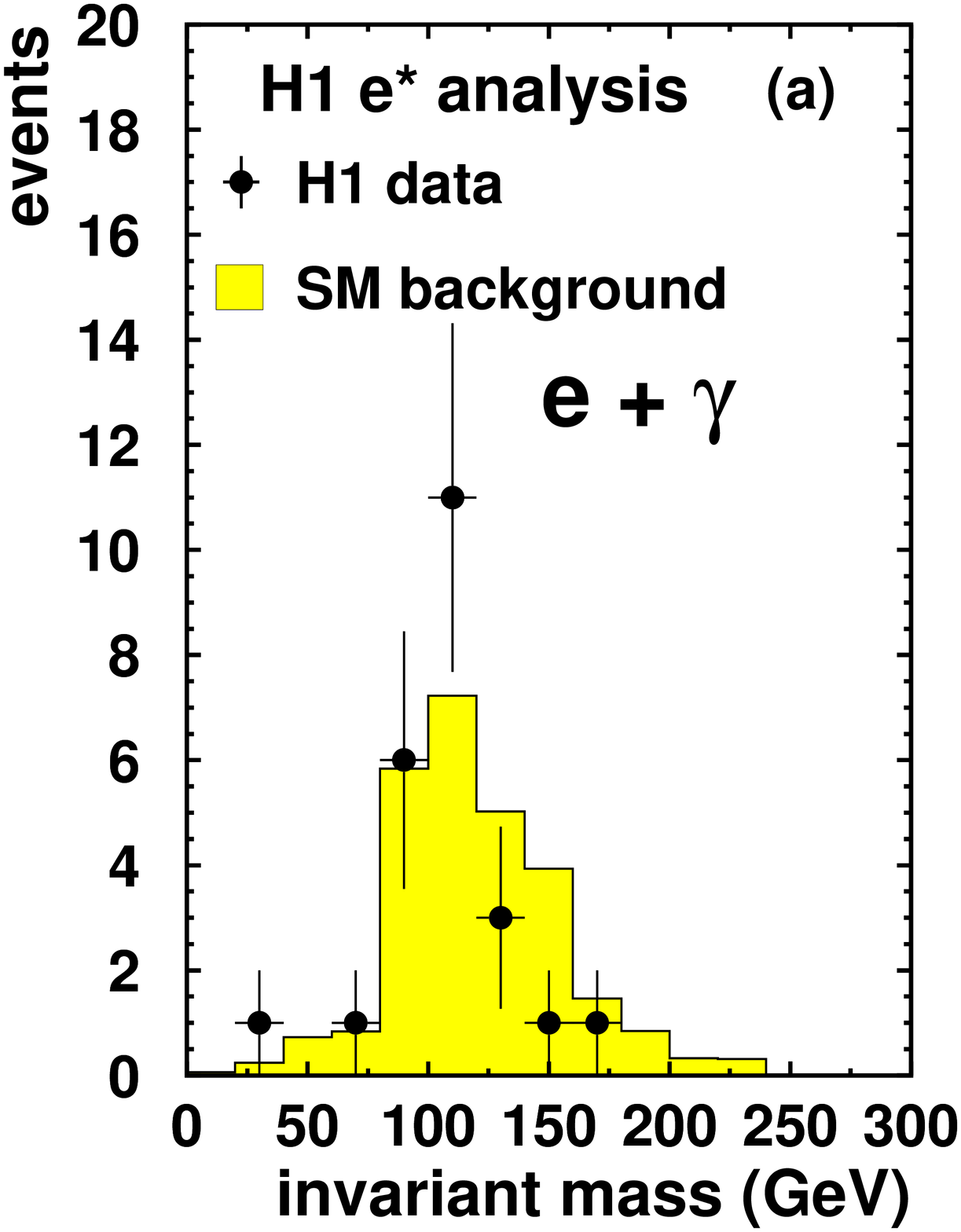} &
\epsfxsize=0.4\textwidth
\epsffile{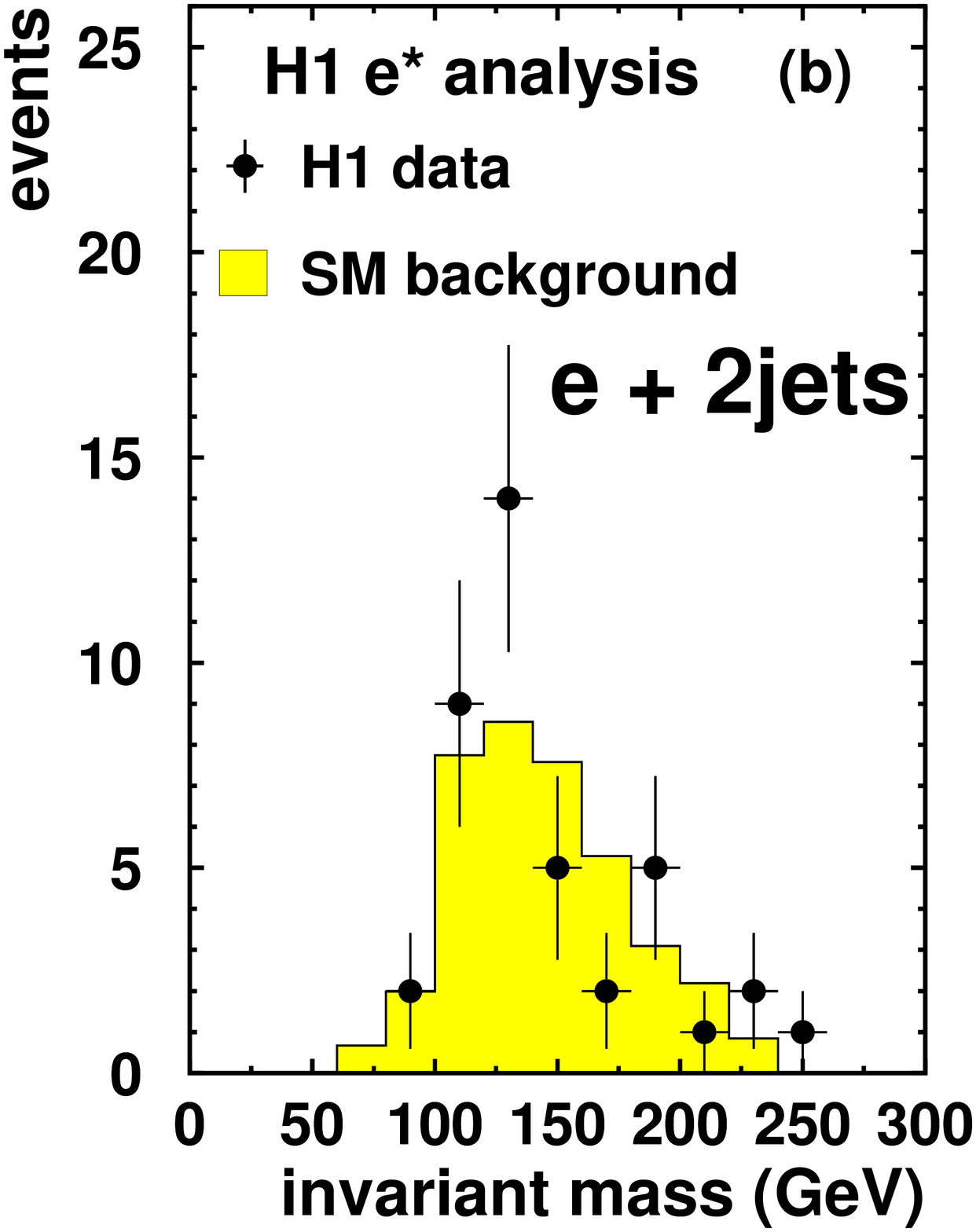} \\
\end{tabular}
\vspace*{-0.5cm}
   \begin{tabular}{p{0.4\textwidth}p{0.5\textwidth}}
   \epsfxsize=0.4\textwidth \epsffile{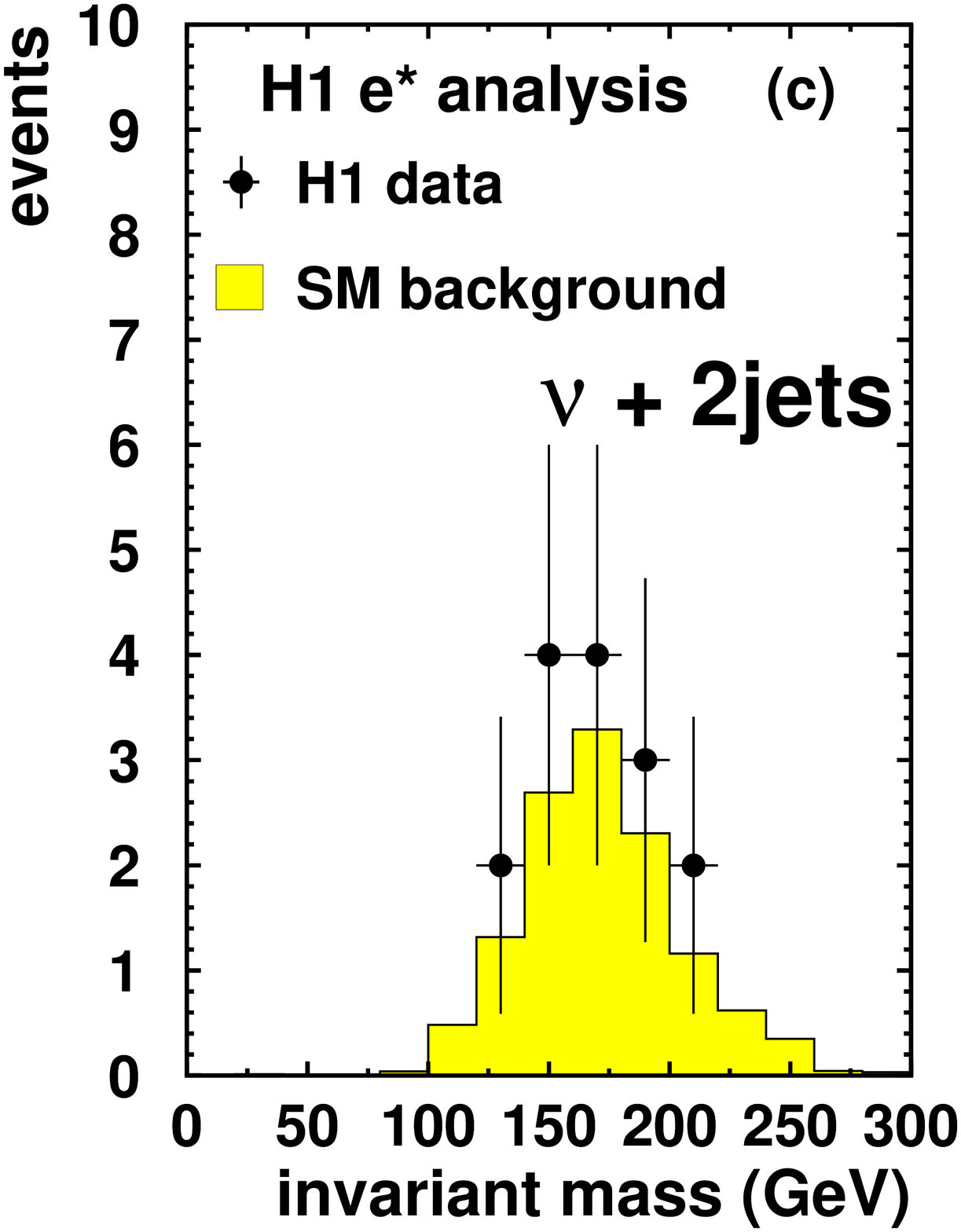} & \vspace*{-7.5cm}

   \caption[]{ \label{fig:masses} {Invariant mass spectra of
candidate (a) $e\gamma$ pairs for the \eega analysis, (b) electron
and two jets  for the $e^* \rightarrow e Z$  analysis and (c)
neutrino and  two jets for the $e^* \rightarrow \nu W$ analysis.
Solid points correspond to the data and the histogram to the total
expectation from different SM processes.}}
   \end{tabular}

\end{center}
\end{figure}

The \eega channel is characterized by two electromagnetic clusters
in the final state. The main sources of background are the WAB
process, NC DIS with photon radiation or a high energy $\pi^0$ in
a jet and the production of electron pairs via $\gamma \gamma$
fusion. Candidate events are selected with two electromagnetic
clusters in the LAr calorimeter of transverse energy greater than
20 GeV and 15 GeV, respectively, and with a polar angle between
0.1 and 2.2 radians. The sum of the energies of the two clusters
has to be greater than 100 GeV. If this sum is below 200~GeV, the
background is further suppressed by rejecting events with a total
transverse energy of the two electromagnetic clusters lower than
75 GeV or with more than two tracks spatially associated to one of the
clusters. The numbers of events passing the analysis cuts for the
SM background processes and for the  data  in each of the three
samples are given in Table~\ref{tab:channels}. About half of the
background originates from NC DIS events with most of the
remainder being due to WAB events. The efficiency for selecting
the signal varies from 85\% for an $e^*$ mass of 150 GeV to 72\%
for an $e^*$ mass of 250 GeV. As in all other channels the
efficiencies are derived using samples of 1000 $e^*$ events
generated at different $e^*$ masses. The various sources of
systematic error are discussed later. Distributions of the
invariant mass of the candidate electron-photon pairs of the three
data samples together and for the SM expectation are shown in
Fig.~\ref{fig:masses}a.

\begin{table}[hhh]
\begin{center}

   \begin{tabular}{|c|c|c|c|c|c|c|}
\hline
\multicolumn{1}{|c}{Sample}   & \multicolumn{2}{|c}{$e^+p\  $ $820\,\mathrm{GeV}$} & \multicolumn{2}{|c}{$e^-p\  $ $920\,\mathrm{GeV}$} & \multicolumn{2}{|c|}{$e^+p\  $ $920\,\mathrm{GeV}$} \\
\hline
  Channel & Data  & SM background  & Data  & SM background &Data  & SM background  \\
\hline
\hline
\eega & 8 & 7.2 $\pm$ 1.0 $\pm$ 0.1 &4 & 4.0 $\pm$ 0.7 $\pm$ 0.2 & 12 & 15.6 $\pm$ 1.7 $\pm$ 0.4 \\
\eezqq & 6 & 7.1 $\pm$ 2.1$\pm$ 2.8 & 4 & 5.6 $\pm$ 0.4 $\pm$ 1.2 & 31 & 25.3 $\pm$ 1.9 $\pm$ 5.5 \\
\enwqq & 2 & 2.4 $\pm$ 0.2 $\pm$ 0.7 &  5 & 3.9 $\pm$ 0.2 $\pm$ 0.7 & 8 & 6.1 $\pm$ 0.4 $\pm$ 1.5 \\
\hline
   \end{tabular}
\caption{Number of candidate events observed in the three decay
channels with the corresponding SM expectation and the
uncertainties on the expectation (statistical and systematic
error).}

\label{tab:channels}

\end{center}
\end{table}

The \eezqq channel is characterized by an electromagnetic cluster
with an associated track and two high transverse energy jets. The
analysis for this channel uses a sample of events with at least
two jets with a transverse energy above 17~GeV and 16~GeV,
respectively, and an electron candidate with a transverse energy
$E_t^{e} > 20$~GeV. These two jets and the electron must have
polar angles smaller than 2.2 radians. Furthermore, to avoid
possible double counting of events from the \eega\ channel, events
with two electromagnetic clusters with transverse energies above
10~GeV and a total energy of the two clusters greater than 100~GeV
are removed. The main SM contribution is NC DIS as photoproduction
events do not yield a significant rate of electron candidates with
large $E_t^{e}$. For 20~GeV $< E_t^{e} <$ 65~GeV, a cut is made on
the electron polar angle. This ranges from $\theta_e < 1.35 $ for
$E_t^{e} = 20$~GeV to $\theta_e < 2.2$~radians for $E_t^{e} =
65$~GeV and depends linearly on $E_t^{e}$. The dijet invariant
mass has to be in the range $- 15< \md  - M_Z < 7$~GeV.  If
there are more than two jets, the pair with invariant mass closest
to the nominal $Z$ boson mass is chosen as the $Z$ candidate. The
two jets chosen are ordered such that $E_t^{jet1} > E_t^{jet2}$.
In many SM events the direction of jet~2 is close to the proton
direction. To ensure that this jet is well measured, an additional
cut on its polar angle, $\theta^{jet2}> 0.2$~radians, is applied
if its transverse momentum is lower than 30 GeV. For an electron
transverse energy 65~GeV~$<~E_t^{e}~<$~85~GeV two jets with an
invariant mass $\md > M_Z - 30$~GeV are required. At very high
transverse energy, $E_t^{e} >$ 85~GeV, the contribution from NC
DIS is very low and no further cuts on $\md$ are needed. The
number of events which remain in the data after these cuts are
summarized in Table~\ref{tab:channels} and compared with the
expected SM contribution (mostly NC DIS events). The efficiency
for selecting the signal varies from 44\% for an $e^*$ mass of
150~GeV to 62\% for an $e^*$ mass of 250~GeV. Distributions of the
invariant mass of the electron and the two jets are shown in
Fig.~\ref{fig:masses}b for data and for the SM expectation.

The \enwqq channel is characterized by two jets and missing
transverse energy \etm . The main background originates from
CC~DIS events with a moderate contribution from photoproduction,
whereas the NC~DIS contribution is suppressed for large \etm . The
analysis starts from a sample of events with at least two jets
with transverse energies above 17 GeV and 16 GeV, missing
transverse energy $\etm~>$~20~GeV and no isolated electromagnetic
cluster with transverse energy above 10 GeV. The jets must have a
polar angle below 2.2 radians. Jets in  which the most  energetic
track enters the boundary region between two calorimeter modules
and central jets ($\theta >$~0.5 radians) are required to contain
more than two tracks. This cut removes NC DIS events in which the
scattered electron is misidentified as a jet. Only events with $S
= \frac{V_{ap}}{V_{p}} <$ 0.1 are accepted, where $V_{ap}$ and
$V_{p}$ are, respectively, the projections of the transverse
energy flow antiparallel and parallel to the transverse momentum
vector of the hadronic system. This cut rejects $ \gamma p$
background for which $S$ is close to 1, whereas for the signal $S$
is close to 0. At very high missing transverse energy,
$\etm~>$~65~GeV, the background is low and no further cuts are
applied. The dijet-pair with invariant mass closest to the nominal
$W$ boson mass is chosen as the $W$ candidate provided its mass is
in the range $-15$~GeV~$ < M_W - \md < 15 $~GeV. The number of
events which remain in the data after these cuts is summarized in
Table~\ref{tab:channels}. Also given is the expected SM
contribution (CC DIS events) for each sample. The efficiency for
selecting the signal varies from 30\% for an $e^*$ mass of 150 GeV
to 52\% for an $e^*$ mass of 250 GeV. Distributions of the
invariant mass of the reconstructed neutrino and the two jets are
shown in Fig.~\ref{fig:masses}c for data and for the SM
expectation.

Contributions to the systematic error of the SM expectation come
from the uncertainty on the absolute energy scale of the
calorimeter and missing higher order corrections in the event
generators which are used for the background estimation. The
uncertainties of the electromagnetic energy scale vary from 0.7\%
in the central part of the detector to 3\% in the forward region.
For the hadronic energy scale an uncertainty of 4\% is assigned.
For the \enwqq and \eezqq channels the SM expectation is varied by
15\% to account for differences observed in particular in two jet
production between perturbative calculations of order
$O(\alpha_s^2)$~\cite{Carli:1998zr,bate,tampere157} and the parton
shower approach. The statistical error of the Monte Carlo samples
is taken into account as a systematic error on the efficiencies.
Finally, the luminosity measurement leads to a normalization
uncertainty of 1.5\%.

\begin{figure}[htbp]
\begin{center}
 \begin{tabular}{p{0.4\textwidth}p{0.6\textwidth}}
    \vspace*{-9.3cm}
      \caption[]{ \label{fig:sigbrH1}
{Upper limits at $95 \%$ confidence level on the product of the
production cross-section $\sigma$ and the decay branching ratio BR
for excited electrons $e^*$ in the various electroweak decay
channels, \eega (dashed line), \eezqq (dotted-dashed line) and
\enwqq (dotted line) as a function of the excited electron mass.
The signal efficiencies used to compute these limits have been
determined with events generated under the assumption $f=+f'$.}}
      &
      \mbox{\epsfxsize=0.55\textwidth
       \epsffile{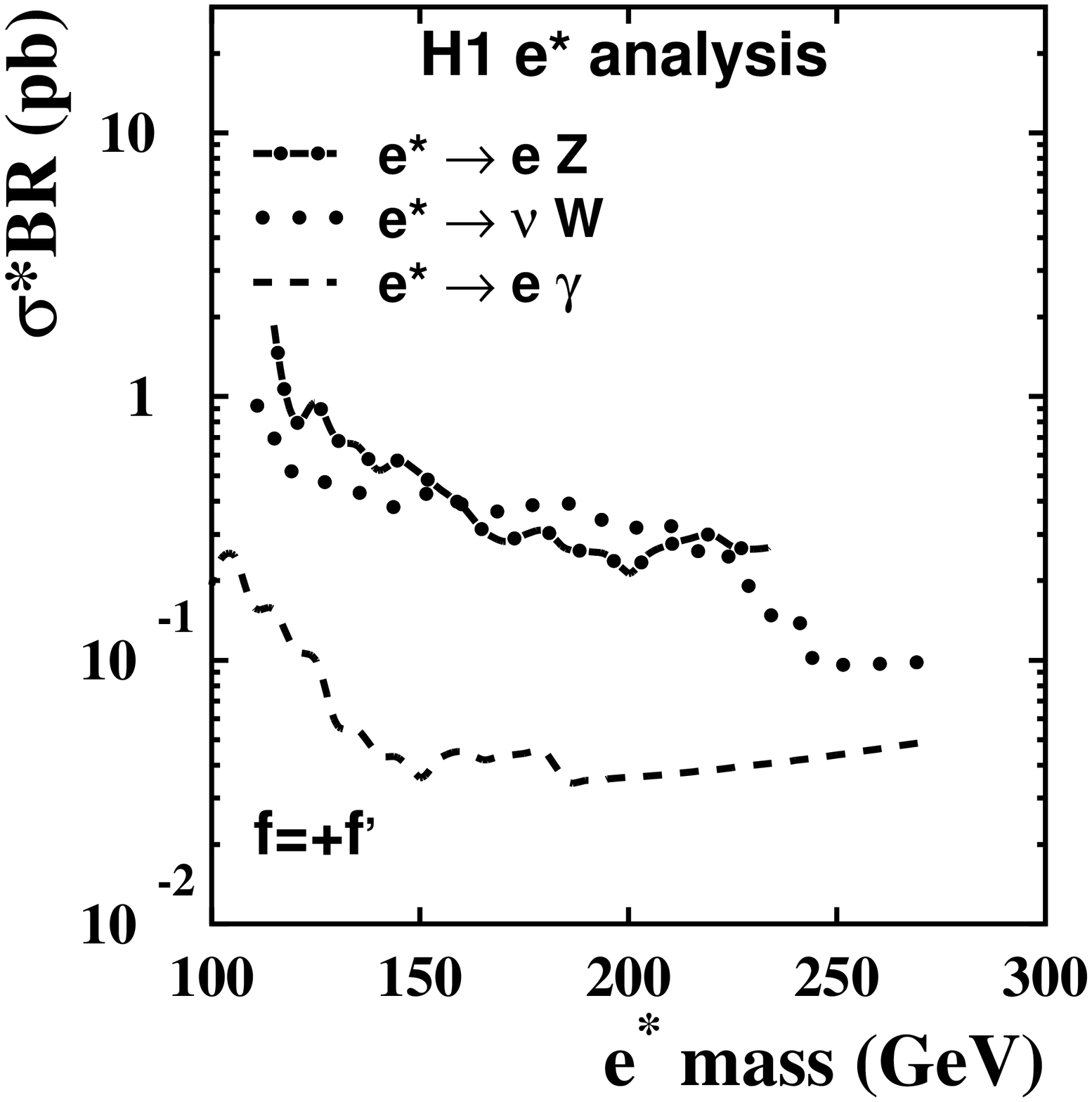}}
\end{tabular}
\end{center}
\end{figure}

\begin{figure}[hhhh]
\begin{center}

\hspace*{-0.18cm}\begin{tabular}{cc}

\epsfxsize=0.52\textwidth
 \epsffile{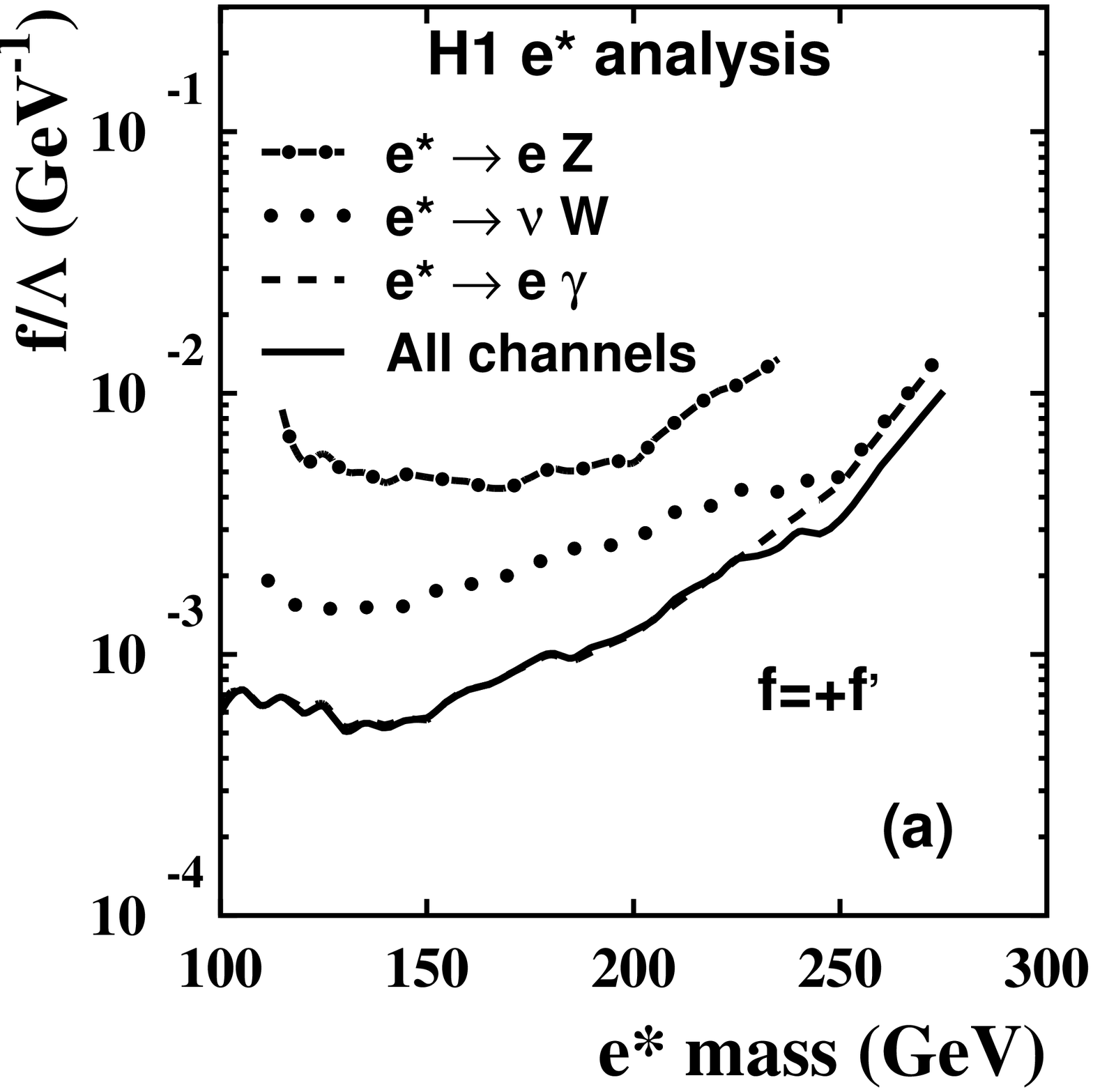} &
\epsfxsize=0.52\textwidth

\hspace*{-0.3cm}\epsffile{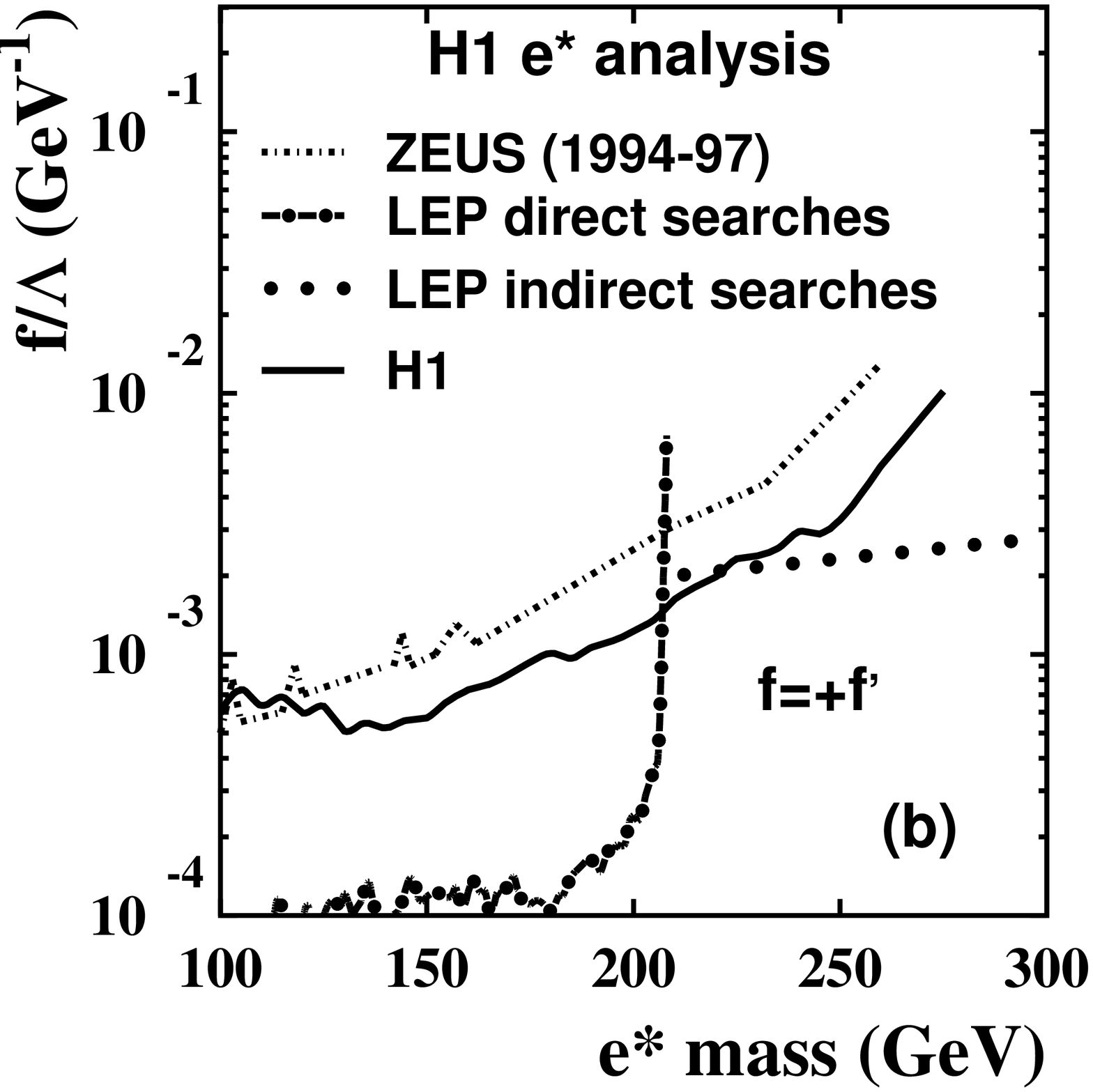} \\
\end{tabular}

\caption{Exclusion limits on the coupling $f/\Lambda$ at $95 \%$
confidence level as a function of the mass of excited electrons
with the assumption $f=+f'$. ({a}) Limit for each decay channel
\eega (dashed line), \eezqq (thick dotted-dashed line), \enwqq
(dotted line) and for the combination of the three channels (full
line). It must be noted that a part of the data included in the
present result were used to obtain the previous H1
limit~\cite{Adloff:2000gv}. ({b}) Comparison of this analysis with
ZEUS results ~\cite{ZEUS} (dashed line) and LEP 2 results on
direct searches~\cite{OPAL} (dotted-dashed line) and on indirect
searches~\cite{Abreu:1998jw} (dotted line). } \label{fig:folH1A}

\end{center}
\end{figure}

In all three search channels the numbers of observed and expected
events are in good agreement. Upper limits on the cross section
and on the coupling $f/\Lambda$ are thus derived as a function of
the $e^*$ mass at 95\% confidence level as described in
\cite{Adloff:2000gv} following the Bayesian
approach~\cite{Barnett:1996hr,Helene:1984ph}. For a given $e^*$
mass, the limits are obtained by counting the number of observed
and expected events in a mass interval that varies with the width
of the expected signal. At $M_{e^*}=150$~GeV, a width of the mass
interval of 30~GeV is chosen for the \eega\  decay mode and 60~GeV
is chosen for the decay channels with two jets. Systematic
uncertainties are taken into account as in~\cite{Adloff:2000gv}.
The limits on the product of the $e^*$ production cross-section
and the decay branching ratio are shown in Fig.~\ref{fig:sigbrH1}.
As stated in the introduction, most experiments give $f/\Lambda$
limits under the assumptions $f=+f'$ and $f=-f'$. The H1 limits
for each decay channel and after combination of all decay channels
are given as a function of the $e^*$ mass in
Fig.~\ref{fig:folH1A}a, for the assumption $f=+f'$. With this
hypothesis the main contribution comes from the \eega \  channel.
The values of the limits for $f/\Lambda$ vary between $5 \times
10^{-4} $ and  $ 10^{-2}\,\mathrm{GeV^{-1}}$ for an $e^*$ mass
ranging from $130\,\mathrm{GeV}$ to $275\,\mathrm{GeV}$. These
results improve significantly the previously published H1 limits
for $e^+p$~\cite{Adloff:2000gv} collisions.

The LEP experiments~\cite{OPAL,Abreu:1998jw} and the ZEUS
collaboration~\cite{ZEUS} have also reported on excited electron
searches. Their limits are shown in Fig.~\ref{fig:folH1A}b. The
LEP~2 experiments have shown results in two mass domains. In
direct searches for excited electrons limits up to a mass of about
$200\,\mathrm{GeV}$ are given. Above $200\,\mathrm{GeV}$ their
results are derived from  indirect searches only. The H1 limit
extends the excluded region to higher masses than reached in
previous direct searches.

\begin{figure}[htbp]
\begin{center}
 \begin{tabular}{p{0.3\textwidth}p{0.7\textwidth}}
    \vspace*{-7cm}
      \caption[]{ \label{fig:folH1B}
{Exclusion limits on the coupling $f/\Lambda$ at $95 \%$
confidence level as a function of the ratio $f'/f$ for three
different masses of the~$e^*$~: 150~GeV (full line), 200~GeV
(dotted line) and  250~GeV (dashed line). }}
      &
      \mbox{\epsfxsize=0.55\textwidth
       \epsffile{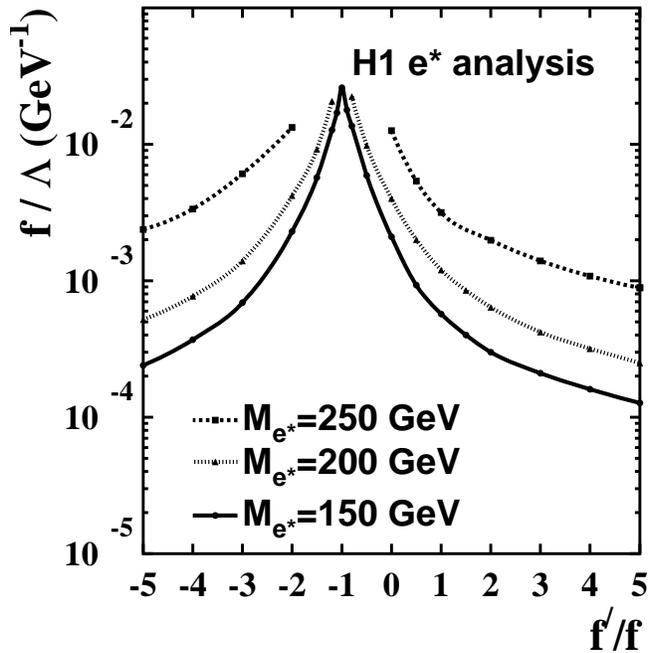}}
\end{tabular}
\end{center}
\end{figure}

More generally, limits on $f/\Lambda$ as a function of $f'/f$ are
shown in Fig.~\ref{fig:folH1B} for three $e^*$ masses ($150$,
$200$ and  $250\,\mathrm{GeV}$). It is worth noting that excited
electrons have vanishing electromagnetic coupling for $f=-f'$. In
this case the $e^*$ is produced through pure $Z$ boson exchange.
As a consequence the production cross-section for excited
electrons at HERA is much smaller in the $f=-f'$ case than in the
$f=+f'$ case. For $e^*$ masses between 150 and 250 GeV the ratio
of the cross-sections for $f=+f'$ and $f=-f'$ varies between 170
and 900. For high $e^{*}$ masses and some values of the couplings,
no limits are given because the natural width of the $e^*$ would
become extremely large.

In summary, a search for excited electron production was performed
using all the $e^+ p$ and $e^- p$ data accumulated by H1 between
1994 and 2000. No indication of a signal was found. New limits
have been established as a function of the couplings and the
excited electron mass for the conventional relationship between
the couplings $f=+f'$. The dependence of the $f/\Lambda$ limit on
the ratio $f'/f$ has been shown for the first time at HERA. The
data presented here restrict excited electrons to higher mass
values than has been possible previously in direct searches.



We are grateful to the HERA machine group whose outstanding
efforts have made and continue to make this experiment possible.
We thank the engineers and technicians for their work in
constructing and now maintaining the H1 detector, our funding
agencies for financial support, the DESY technical staff for
continual assistance and the DESY directorate for the hospitality
which they extend to the non-DESY members of the collaboration.


\end{document}